\documentstyle[11pt]{article} 
\def\baselinestretch{1.3}
\begin{document}
\begin{flushright}
hep-th/0204242
\end{flushright}

\begin{center}
{\Large\bf  Does a Randall-Sundrum scenario create the illusion of 
a torsion-free universe?}\\[20mm]
Biswarup Mukhopadhyaya\footnote{E-mail: biswarup@mri.ernet.in},
Somasri Sen \footnote{E-mail: somasri@mri.ernet.in}\\
{\em Harish-Chandra Research Institute,\\
Chhatnag Road, Jhusi, Allahabad - 211 019, India} 

Soumitra SenGupta \footnote{E-mail: tpssg@mahendra.iacs.res.in} \\
{\em Department of Theoretical Physics, Indian Association for the 
Cultivation of Science,\\
Calcutta - 700 032, India}\\[20mm] 
\end{center}

\vspace{0.5cm}
\vspace{0.5cm}

\begin{abstract}
We consider spacetime with torsion in a Randall-Sundrum (RS) scenario
where torsion, identified with the rank-2 Kalb-Ramond field, exists in
the bulk together with gravity. While the interactions of
both graviton and torsion in the bulk are controlled by the Planck 
mass, an additional exponential suppression comes for the torsion zero-mode on
the visible brane. This may serve as a natural explanation of why the effect 
of torsion is so much weaker than that of curvature on the brane. 
 The massive torsion
modes, on the other hand, are correlated with the corresponding 
gravitonic modes and may be detectable in TeV-scale experiments.

\end{abstract}

\vskip 1 true cm

\newpage
\setcounter{footnote}{0}

\def\baselinestretch{1.8}

Theories with large compact extra dimensions have gained considerable 
attention in recent times, primarily because of their role in solving
the naturalness problem. A high point of such theories is the prediction
of TeV-scale observable effects of the Kaluza-Klein modes of the gravitational
field which is assumed to exist in the `bulk'. Such speculation can be
broadly classified into two schools of thought, based essentially on the
approaches of Arkani-Hamed, Dimopoulos and Dvali (ADD)\cite{add} on one hand, 
and Randall and Sundrum (RS)\cite{rs} on the other. In both types of 
models, all
visible matter (i.e. the content of the standard model (SM) of particle
interactions) is supposed to be confined to a `3-brane' on which the 
projections of the bulk gravity gives rise to Kaluza-Klein modes. The 
spacings of these modes and their interactions with the SM fields 
are determined by specifics of the model.

In various extensions of the above models, implications of other types of
bulk fields, such as scalars, gauge fields and fermions, 
have been explored\cite{gw}-\cite{addf}. In this note, 
we examine what happens if bulk spacetime 
in an RS picture is endowed with both curvature and torsion, a possibility
that is motivated from string theory\cite{gsw}. 
From such an assumption, we try to explain why, sitting on the visible brane, 
one might feel the presence of curvature but not of torsion, although both 
might have originally (i.e. at the bulk level) been on the same footing.

It has been an old suggestion to modify theories of gravity by incorporating
torsion in space-time along with curvature. The most straightforward way
of including torsion is to add an antisymmetric component to the 
connection $\Gamma_{\mu\nu}^\alpha$. This is the essence of 
the so-called Einstein-Cartan type of theories\cite{hehl}.

Once torsion enters into the theory in the above manner, it can couple
with all matter fields with spin. It can be easily seen that such interaction
terms in general are of dimension 5, and are suppressed by the Planck mass 
($M_P$), much in the same way as in the case of gravitonic couplings. 
Efforts have been on to constrain the torsion field and its coupling strength 
from a variety of considerations such as atomic energy level splitting
\cite{lae} and the phenomena of optical activity in radiation from distant 
galactic sources\cite{sps,spss}. 
However, it is not clearly understood from any {\em fundamental 
theoretical consideration} whether the coupling of torsion with visible matter
should be different from that of curvature, and if so, why. This is
precisely the question we address here, within the framework of
an RS theory.

In its minimal version the RS scenario, defined in 5-dimensions\cite{rs}, 
is characterised by the background metric

\begin{equation}
ds^2=e^{-2\sigma}\eta_{\mu\nu}dx^{\mu}dx^{\nu}+r_c^2d\phi^2
\end{equation}

\noindent
with $\eta_{\mu\nu}~=~(-,+,+,+)$, and $\sigma~=~k{r_c}|\phi|$. 
$r_c$ is the compactification
radius for the fifth dimension, and $k$ is on the order of the
higher dimensional Planck mass $M$. The extra dimension, characterized
by a variable $\phi$ ranging from $-\pi$ to $+\pi$, forms
an $S_{1}/Z_{2}$ orbifold. The standard model fields reside at
$\phi~=~\pi$ while gravity peaks at $\phi~=~0$. The dimensional parameters 
defined above are related to the  4-dimensional Planck scale $M_P$
through the relation

\begin{equation}
M_P^2=\frac{M^3}{k}[1-e^{-2kr_c\pi}]
\end{equation}

Clearly, $M_P$, $M$ and $k$ are all of the same order of magnitude. 
For $k r_c~\simeq~12$ the exponential factor (frequently referred to as the
`warp factor') produces TeV scale mass parameters 
(of the form $m~=~Me^{-kr_{_c}\pi}$) from the Planck scale
when one considers projections on the `standard model' brane. Thus the 
hierarchy between the Planck and TeV scales can be accommodated without
the need of fine-tuning.

It is well-known that the likely source of torsion is some matter field(s)
with spin, just as curvature is associated with mass/energy. Attempts 
have been made in some earlier works\cite{leb1,leb2} to relate torsion with 
fermion fields 
residing either on the brane or in the bulk. Here we take the standpoint that,
being as much a characteristic of spacetime as curvature, it is natural 
for torsion to coexist with gravity in the bulk. We have earlier performed
some analyses in this line in the context of an ADD model\cite{bss}.

In the scenario adopted by us, the source of torsion is taken to be the
rank-2 antisymmetric Kalb-Ramond (KR) field $B_{MN}$ which arises as a
massless mode in heterotic string theories\cite{gsw}. 
To understand the above statement, let us recall that the 
low energy effective action for the gravity and  Electromagnetic sectors 
in D dimensions is given by

\begin{equation}
S = \int~ d^{D}x \sqrt{-G} ~\left[~R(G) ~-~
    \frac{1}{4} F_{MN}F^{MN} ~+~
    \frac{3}{2} H_{MNL}H^{MNL} ~\right]
\end{equation}

It has been shown earlier \cite{pmssg} that an action of the form

\begin{equation}
S = \int~ d^{D}x \sqrt{-G} ~\left[~R (G,T) ~-~
    \frac{1}{4} F_{MN} F^{MN} ~-~
    \frac{1}{2}H_{MNL} H^{MNL} ~+~
     T_{MNL}H^{MNL}\right]
\end{equation}

\noindent
reproduces the low energy string effective action if one eliminates the 
torsion field 
$ T_{MNL}$ ( which is an auxiliary field) by using the equation of motion
$ T_{MNL} ~=~ H_{MNL}$.

Thus torsion can be identified 
with the rank-3 antisymmetric field strength tensor $H_{MNL}$ 
which in turn is related to the KR field $B_{MN}$\cite{kr} as

\begin{equation}
H_{MNL} = \partial_{[M}B_{NL]}
\end{equation}

\noindent
with each Latin index running from $0$ to $4$. (Greek indices, on 
the other hand, run from 0 to 3.) Furthermore, we use the KR gauge fixing 
conditions to set $B_{4\mu}~=~0$. Therefore, the only non-vanishing 
KR field components correspond to the brane indices. These components,
of course, are functions of both compact and non-compact co-ordinates.

The 5-dimensional action for the curvature-torsion sector in this case is

\begin{equation}
{\cal S}_G=\int d^4x\int d\phi~\sqrt{-G}~2~M^3~ R(G,H)
\end{equation}

\noindent
where $G_{MN}$ is given by equation (1) and $R(G,H)$ is the scalar 
curvature constructed from the modified affine connection:

\begin{equation}
{\bar{\Gamma}^K}_{NL}={{\Gamma}^K}_{NL}-\frac{1}{M^{\frac{3}{2}}}{H_{NL}}^K
\end{equation}

Clearly, the action can be decomposed into two independent parts- one 
consisting of pure curvature, and the other, of torsion:

\begin{equation}
{\cal S}_G=\int d^4x\int d\phi~\sqrt{-G}~2[M^3~ R(G)-H_{MNL}~H^{MNL}]
\end{equation}

\noindent
with $H_{MNL}$ related to the Kalb-Ramond field $B_{NL}$ as in equation (3).

Thus the 5-dimensional action corresponding to the Kalb-Ramond field,
upto a dimensionless multiplicative constant, is given by

\begin{equation}
{\cal S}_H=\int d^4x\int d\phi~\sqrt{-G}~H_{MNL}~H^{MNL}
\end{equation}

Using the explicit form of the RS metric, and remembering that $B_{4\mu}~=~0$,
we have

\begin{equation}
{\cal S}_H=\int d^4x\int d\phi~r_c~e^{2\sigma(\phi)}~[\eta^{\mu\alpha}\eta^{\nu\beta}
\eta^{\lambda\gamma}H_{\mu\nu\lambda}H_{\alpha\beta\gamma}-\frac{3}{r_c^2}e^{-2\sigma(\phi)} \eta^{\mu\alpha}\eta^{\nu\beta}B_{\mu\nu}\partial_\phi^2
~B_{\alpha\beta}]
\end{equation}

Next, we consider Kaluza-Klein decomposition for the Kalb-Ramond field:

\begin{equation}
B_{\mu\nu}(x,\phi)=\sum_{n=0}^{\infty}~B^n_{\mu\nu}(x)\frac{\chi^n(\phi)}{\sqrt{r_c}}
\end{equation}

In terms of the four-dimensional projections $B^{n}_{\mu\nu}$,
an effective action of the form

\begin{equation}
{\cal S}_H=\int d^4x~\sum_{n=0}^{\infty}~
[\eta^{\mu\alpha}\eta^{\nu\beta}
\eta^{\lambda\gamma}H^n_{\mu\nu\lambda}H^n_{\alpha\beta\gamma}+3m_n^2\eta^{\mu\alpha}\eta^{\nu\beta}B^n_{\mu\nu}B^n_{\alpha\beta}]
\end{equation}

\noindent
can be obtained provided

\begin{equation}
-\frac{1}{r_c^2}\frac{d^2\chi^n}{d\phi^2}=m_n^2\chi^n e^{2\sigma}
\end{equation}

\noindent
and subject to the orthonormality condition

\begin{equation}
\int e^{2\sigma(\phi)} \chi^m(\phi)\chi^n(\phi)d\phi=\delta_{mn}
\end{equation}  

\noindent
where $H^n_{\mu\nu\lambda}~=~\partial_{[\mu}B^n_{\nu\lambda ]}$ and 
$\sqrt{3}m_n$ gives the mass of the $n$th mode.
In terms of $z_n~=~{\frac{m_n}{k}}e^{\sigma(\phi)}$, equation (11) 
can be recast in the form

\begin{equation}
\left [z_n^2\frac{d^2}{dz_n^2}+z_n\frac{d}{dz_n}+ z_n^2\right]\chi^n=0 
\end{equation}  

The above equation admits of the following solution:

\begin{equation}
\chi^n=\frac{1}{N_n}\left[J_0(z_n)+\alpha_n Y_0(z_n)\right]
\end{equation}  

\noindent
where $J_{0}(z_n)$ and $Y_{0}(z_n)$ are respectively Bessel and Neumann
functions of order zero. $\alpha_n$ as well as $m_n$ can be determined
from the continuity conditions for the derivative of $\chi_n$ at 
$\phi~=~0$ and $\pi$, which are dictated by self-adjointness of the
left-hand side of equation (11). On using the fact that $e^{kr_c\pi}>>1$
and the  mass values $m_n$ on the brane should be on the order 
of the TeV scale $(<<k)$, 
we obtain from the continuity condition  at $\phi=0$

\begin{equation}
\alpha_n \simeq x_n e^{-2kr_c\pi} 
\end{equation}

\noindent
with $x_n~=~z_n(\pi)$. The boundary condition at $\phi=\pi$ gives 

\begin{equation}
J_1(x_n)\simeq \frac{\pi}{2}x_n e^{-2kr_c\pi} 
\end{equation}

Since the right-hand side of the above equation is very small, the roots
can be closely approximated to the zeros of $J_{1}(x_n)$. These roots of 
$J_1$ give $m_n$ on the TeV range, as expected initially.

Since $x_n~\simeq~1$,  from equation (15) $\alpha_{n}$ becomes $ << 1$. 
The normalization condition yields

\begin{equation}
N_n=\frac{\pi}{2\sqrt{kr_c}}x_n e^{-kr_c\pi} 
\end{equation}

\noindent
Thus the final solution for the massive modes turns out to be

\begin{equation}
\chi^{n}(z_n) = \frac{2\sqrt{kr_c}}{\pi x_n}e^{kr_c\pi}~J_0(z_n) 
\end{equation}

At this point it is useful to compare the solutions with those for 
bulk gravitons\cite{hrd2} and gauge fields\cite{hrd1}. First, here 
the massive solutions
are governed by zeroth-order Bessel functions, as against second\cite{hrd2}
 and first order\cite{hrd1} ones in the two other cases. Furthermore, 
a comparison 
with the above references shows that whereas a massive gravitonic mode
contains the same exponential enhancement factor as that in
equation (18), it is absent in the case of bulk gauge fields. This 
difference can be attributed essentially to the tensorial structures
of the different types of bulk fields 
 as well as to the characteristic forms 
of the 4-dimensional effective actions into which the theory must reduce in 
the different cases.

\begin{center}
$$
\begin{array}{|c|c c c c|}
\hline
n  & 1  & 2  & 3  & 4  \\
\hline
m_n^{grav}~(TeV) & 1.66 & 3.04 & 4.40 & 5.77 \\
\hline
m_n^{tor}~(TeV)  & 2.87 & 5.26 & 7.62 & 9.99 \\
\hline
\end{array}
$$ 
\end{center} 


 Table 1: The masses of a few low-lying gravitonic modes vis-a-vis
     the massive KR modes for $kr_c=12$ and $k=10^{19}$Gev.

As table 1 shows, the mass spectrum here is correlated
with the masses of the gravitonic modes. This is because the masses in
both cases are effectively given, upto an overall factor of $\sqrt{3}$, 
by the zeros of $J_{1}(x_n)$ (as we have already noticed, the right-hand side 
of equation (16) is negligibly small). Thus the scenario proposed here
has an added element of predictability as far as the graviton and torsion KK 
modes are concerned. The  $B^n_{\mu\nu}$ spacings ($n=1$ onwards) are 
just scaled with respect to the gravitonic modes by a factor of
$\sqrt{3}$ to a close approximation, and therefore the low-lying
states in the spectrum should be within the reach of TeV-scale collider
experiments.

However, a  more drastic difference is noticed when we consider the
massless mode. In this case the solution to (11) turns out to be

\begin{equation}
\chi^0(\phi) = c_1 |\phi| + c_2
\end{equation}

The condition of 
self-adjointness leaves the scope of only a constant 
solution. Using the normalization condition, one obtains

\begin{equation}
\chi^0 = \sqrt{k r_c} e^{-k r_c \pi}
\end{equation}

Thus, in contrast to the other types of bulk fields 
mentioned above, the zero mode $\chi^{0}$ exhibits 
{\em a suppression by a large exponential factor.} This causes the massless KR
mode to be severely  suppressed on the visible brane, granting a practically
imperceptible presence to torsion.

This can be seen more clearly if we consider the coupling of torsion to matter 
fields on the visible brane. Let us for example consider the interaction
with spin-1/2 fields\cite{bs}. Starting with a 5 dimensional action 
and remembering 
that the fermion and all its interactions are confined to the brane at 
$\phi=\pi$, the fermionic action in terms of the modified affine connection 
is given by:

\begin{equation}
{\cal S}_\psi=i\int d^4x \int d\phi~[{\it det}~V]
~\bar{\psi}~[\gamma^{a}v^\mu_a~(\partial_\mu-\frac{i}
{2}G_{LN}\sigma^{ab}v^\nu_a\partial_\mu v^\lambda_b\delta^N_\nu\delta^L_\lambda-G_{AD}\sigma^{ab}v^\beta_a v^\delta_b{\bar{\Gamma}_
{MB}}^A\delta^M_\mu\delta^B_\beta\delta^D_\delta)]~\psi~\delta(\phi-\pi)
\end{equation}

\noindent
where $G_{MN}$ is given by,

\begin{equation}
G_{MN}=v^a_M v^b_N\eta_{ab}
\end{equation}

\noindent
and the vierbein $v^a_\mu$ is given in this case by

\begin{equation}
v^4_4=1;~~v^a_\mu=e^{-\sigma}\delta^a_\mu;~~{\it det}~V=e^{-4\sigma}
\end{equation}

\noindent $a,~b$ etc. being tangent space indices.

Integrating out the compact dimension and using the fact that the
fermion field on the brane is consistently renormalized as 
$\psi\rightarrow e^{3kr_c\pi/2} \psi$, one obtains the effective 
4-dimensional fermion KR interaction as

\begin{equation}
{\cal L}_{\psi\bar{\psi}H}= - \bar{\psi}[i\gamma^{\mu}\sigma^{\nu\lambda}
\left\{\frac{1}{M_P e^{kr_c\pi}}H^0_{\mu\nu\lambda}+\frac{1}{\Lambda_{\pi}}
\frac{2J_0(x_n)}{\pi x_n}\sum_{n=1}^{\infty}H^n_{\mu\nu\lambda}\right\}]\psi
\end{equation}

\noindent
where $H^n_{\mu\nu\lambda}~=~\partial_{[\mu}B^n_{\nu\lambda ]}$ and
$\Lambda_\pi=M_P e^{-kr_c\pi}$.

A rather remarkable fact becomes evident from above. Although gravity
and torsion are treated at par on the bulk, with the Planck mass 
characterizing any dimensional parameter controlling their interactions,
the coupling of the zero-mode torsion field with fermionic matter suffers
an enormous additional suppression via the warp factor when one compactifies 
the extra dimension in the RS scheme. However, the massless graviton 
continues to have interactions driven by $1/M_P$ on the brane. 
In a way, this leads to the conclusion that the experimental signatures
of torsion will continue to be elusive
so long as we are sitting on the brane, with the apparent feeling that
we are living in a torsionless universe.

The massive modes $H^n_{\mu\nu\lambda}$, however, have enhanced coupling with
matter, caused by the usual warp factor. 
The lowest-lying modes are in the
TeV scale and their interaction strength is suppressed by a mass of similar
magnitude. One can hope to see observable effects of these modes through new 
resonances in TeV scale accelerator experiments, or in the helicity flip
of, say, high-energy massive neutrinos interacting with torsion\cite{as}.

As far as the interaction of bulk torsion with standard model gauge fields 
is concerned, very similar effects can be seen. We show this in the context 
of the $U(1)_{EM}$ gauge field. First it has to be remembered
that gauge invariant interactions of this type cannot be obtained
without introducing some amount of nonminimality into the theory. A 
rather convenient way of doing so is to augment the brane components of the 
torsion tensor $H_{MNL}$ with a so-called Chern-Simons term\cite{gsw}:

\begin{equation}
H_{\mu\nu\lambda}(x,\phi=\pi) ~\rightarrow~ 
H_{\mu\nu\lambda}(x,\phi=\pi) 
~+~\frac{1}{M^{\frac{1}{2}}} A_{[\mu}(x) F_{\nu\lambda]}(x)
\end{equation}

Such a term is natural in the context of gauge anomaly cancellation
in a heterotic string theory\cite{gsw}.
On considering the KK modes $B^n_{\mu\nu}$, and appropriately
redefining the electromagnetic field on the brane, one ends up with
interaction terms of the following nature:

\begin{equation}
{\cal L}_{em-H}= - \frac{1}{M_P e^{kr_c\pi}}A^{[\mu}F^{\nu\lambda]}(x)
H^0_{\mu\nu\lambda}(x)
- \frac{1}{\Lambda_{\pi}}\frac{2J_0(x_n)}
{\pi x_n}A^{[\mu}F^{\nu\lambda]}(x)\sum_{n=1}^{\infty}H^n_{\mu\nu\lambda}(x)
\end{equation}

\noindent
where again we notice an additional suppression for the massless modes
and an enhancement for the massive ones, just as in the case of
spin-1/2 particles coupling to torsion. This extreme suppression of
the zero-mode coupling could weaken torsion-induced optical activities, 
as suggested in some recent works\cite{sps,spss}.

The following picture emerges from the above analysis. Torsion, caused by
a KR field, can be postulated
to exist in the bulk in a 5-dimensional RS scenario, together 
with gravity. At that level, only one mass parameter (of the order of the 
Planck scale) controls the actions for both gravity and torsion.  
The compactification  of the fifth dimension gives rise to
a spectrum of Kaluza Klein modes for the KR field, just as in the case 
of gravity. However, the torsion zero
mode on the visible brane has an added suppression through the warp factor
in its interaction with matter fields. This  indicates that it is going to
be nearly impossible to see its trace in any observation performed
on the visible brane. Therefore, we shall continue to have the 
impression of residing in a torsionless universe if this scenario is 
correct. On the other hand, the massive KR spectrum gets correlated 
with the spectrum of graviton, and their signals in TeV scale accelerator 
experiments can make the hypothesis of bulk torsion verifiable.

\noindent
{\bf Acknowledgement:} The work of BM and SSG has been partially supported by
the Board of Research in Nuclear Sciences, Government of India. SSG 
acknowledges the hospitality of Harish-Chandra Research Institute while this 
work was in progress. SS thanks P. Crawford for hospitality at
the University of Lisbon where part of the work was done.

\end{document}